\documentclass{emulateapj}
\usepackage{amssymb}
\usepackage{amsmath}
\usepackage{mathrsfs}
\usepackage{natbib}
\usepackage[colorlinks, linkcolor=blue, urlcolor=blue, citecolor=blue]{hyperref}
\usepackage{array}
\usepackage{float}
\usepackage{graphicx}
\usepackage{subfigure}
\usepackage{rotating}
\usepackage{color}
\usepackage[all]{hypcap}
\usepackage[toc,page]{appendix}

\RequirePackage{lineno}
\slugcomment{}
\journalinfo{}
\makeatletter\def\frontmatter@title@below{\vspace*{-2.63\baselineskip}\vspace*{0.25in}}\makeatother

\begin{document}

\title{Second Repeating FRB 180814.J0422+73: Ten-year \textit{Fermi}-LAT Upper limits and Implications}
\author{Yu-Han Yang\altaffilmark{1,2}, Bin-Bin Zhang\altaffilmark{1,2}, and Bing Zhang\altaffilmark{3}}

\begin{abstract}
The second repeating fast radio burst source, FRB 180814.J0422+73, was detected recently by the CHIME collaboration. We use the ten-year \textit{Fermi} Large Area Telescope (LAT) archival data to place a flux upper limit in the energy range of 100~MeV$-$10~GeV at the position of the source, which is $\sim {1.1\times 10^{-11}}$~erg~cm$^{-2}$~s$^{-1}$ for a six-month time bin on average, and $\sim {2.4\times 10^{-12}}$~erg~cm$^{-2}$~s$^{-1}$ for the entire ten-year time span. For the maximum redshift of $z=0.11$, the ten-year upper limit of luminosity is $\sim {7.3\times 10^{43}}$~erg~s$^{-1}$. We utilize these upper limits to constrain the FRB progenitor and central engine. For the rotation-powered young magnetar model, the upper limits can pose constraints on the allowed parameter space for the initial rotational period and surface magnetic field of the magnetar. We also place significant constraints on the kinetic energy of a relativistic external shock wave, ruling out the possibility that there existed a gamma-ray burst (GRB) beaming towards Earth during the past ten years as the progenitor of the repeater. The case of an off-beam GRB is also constrained if the viewing angle is not much greater than the jet opening angle. All these constraints are more stringent if FRB 180814.J0422+73 is at a closer distance.
\end{abstract}

\keywords{pulsars: general -- radiation mechanisms: non-thermal -- stars: magnetars -- stars: magnetic field}

\affil{\altaffilmark{1}School of Astronomy and Space Science, Nanjing
University, Nanjing 210093, China; bbzhang@nju.edu.cn}
\affil{\altaffilmark{2}Key Laboratory of Modern Astronomy and Astrophysics (Nanjing University), Ministry of Education, China}
\affil{\altaffilmark{3}Department of Physics and Astronomy, University of Nevada, Las Vegas, NV 89154, USA}

\section{Introduction}
Fast radio bursts (FRBs) are GHz-band radio transient sources with typical durations of milliseconds \citep{lorim07, kean12, thor13, spit14, masui15, ravi15, cham16, ravi16, petr16}. So far there are 79 reported FRBs\footnote{\url{http://frbcat.org/}} detected by various radio telescopes \citep{petr16}, among which two are repeating sources. It is possible that all FRBs repeat, but repeating FRBs may form a sub-class of FRBs that have a distinct origin from the one-time FRB events \citep{palaniswamy18, caleb19}. For a long time period, FRB 121102 \citep{spit16, scho16, chat17, marc17, gajj18, zhang18} has been the sole repeating FRB source. The second repeating event, FRB 180814.J0422+73, was recently discovered by the CHIME/FRB team \citep{2019chime}, suggesting that there could be many repeaters. The dispersion measure (DM) of FRB 180814.J0422+73 is about 189~pc~cm$^{-3}$, including the contribution from the disk of the Milky Way, \mbox{87--100~pc~cm$^{-3}$}  \citep{2019chime}. By assuming that all the excess DM comes from the intergalactic medium, the inferred upper limit of redshift is $z\le 0.11$, corresponding to a luminosity distance of $D \le 500 $~Mpc. The best estimate of its J2000 position is R.A.~$04^h22^m22^s$, Dec.~$+73^\circ 40'$\citep{2019chime}.

The physical origin of FRBs is unknown. One particular model invokes a young magnetar as the source \citep[e.g.][]{murase16,lyu16,megz17, kashi17}. The host galaxy of FRB 121102 is similar to that of a long duration GRB or a super-luminous supernova (SLSN) \citep{marc17, tend17}. This led to the suggestion that a long GRB or a SLSN may be the progenitor of the young magnetar that powers the repeating FRBs \citep{megz17,nicholl17,margalit18}. Within such a picture, there could be a GRB or SLSN that preceded the repeating FRBs within the time scale of years\footnote{A more direct connection between GRBs and FRBs was proposed by \cite{zhang14}, but the FRB in that model is a catastrophic event and does not repeat.}.

In our previous work \citep{2017zhang}, we processed the {\it Fermi} Large Area Telescope (LAT) data and presented a flux and luminosity upper limit on the gamma-ray emission from FRB 121102. The non-detection places some constraints on the allowed parameters of the underlying putative magnetar. At the redshift 0.19 of FRB 121102 \citep{tend17}, the constraints on the magnetar parameter space were not tight.
In this paper, we perform a similar analysis to the second repeater FRB 180814.J0422+73 (Sect 2). Since it is at a closer distance (redshift upper limit 0.11), the constraints on the magnetar parameter are tighter (Sect.3.1). Furthermore, we can also use the non-detection results to place a strong constraint on the existence of a long GRB during the past ten years as the FRB progenitor (Sect.3.2). Our results are summarized in Sect. 4.

\section{\textit{Fermi}-LAT observations in the direction of FRB 180814.J0422+73}

\begin{figure}[tbph]
\begin{center}
\includegraphics[width=1 \textwidth,angle=0]{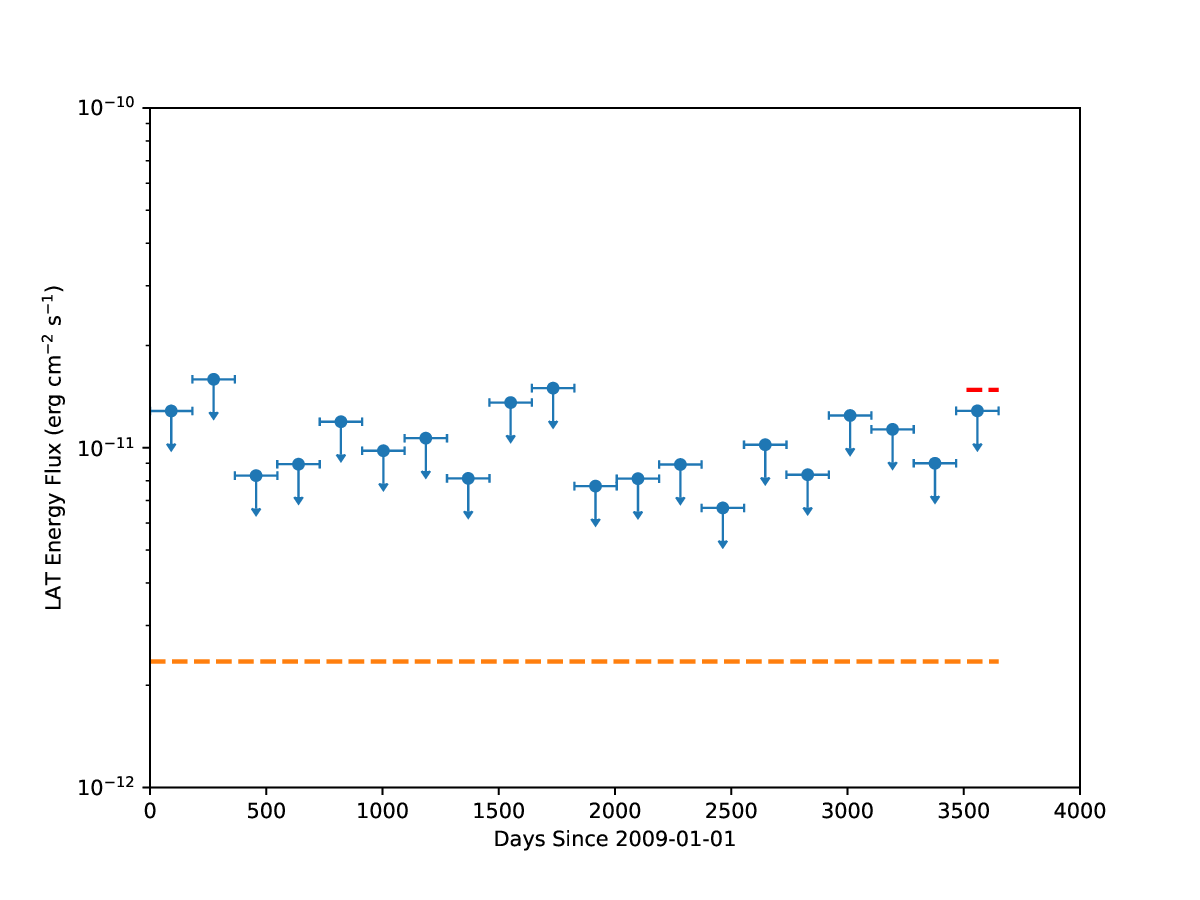}
\end{center}
\caption{ {The flux upper limits of FRB 180814.J0422+73 in the 100MeV$-$10GeV band.} The time range is from January 1, 2009 to December 29, 2018. Each single upper limit is for a half-year bin. The orange dashed line denotes the ten-year upper limit extrapolated from the average value of the 20 half-year upper limits. The red dashed line is the upper limit with a time span between August 14, 2018 and December 29, 2019, which is derived from the upper limit of the last half-year bin.}
\label{fig:lim}
\end{figure}

Thanks to its wide energy range, large FOV, and continuous temporal coverage, \textit{Fermi}-LAT has monitored the direction FRB 180814.J0422+73 for ten years and is an ideal instrument to search for a possible $\gamma$-ray counterpart of the source. We select the energy range of 100~MeV$-$10~GeV and a ten-year time span from 2009 January 1 to 2018 December 29, to extract and process the LAT data. The standard {\em Fermi} Science Tools (v11r5p3) are employed to process the Pass 8 data. We divide the time range into 20 six-month bins, in which no significant (TS $<$ 25) source is detected. The Test Statistic ${\rm TS}=2(\mathcal{L}_1 -\mathcal{L}_0) $, where $\mathcal{L}_1$ and $\mathcal{L}_0$ are the log-likelihood with and without the source which is used in the unbinned maximum likelihood algorithm \citep{abdo10a}. TS $<$ 25 corresponds to $\lesssim 4\sigma $ statistical significance \citep{abdo10b}. Therefore, we utilize the ``integral" method to calculate the upper limit of the photon flux at the 95$\%$ confidence level \citep{feldman98, roe99}. A power-law spectrum model with the average photon index $\sim -2$ is assumed. The upper limits of energy flux for all the time bins can be derived, and the isotropic luminosity upper limits can be calculated by assuming the maximum redshift of 0.11. The ten-year upper limit is extrapolated from the average value of the total 20 half-year upper limits using the relation $L_{\rm lim} (t) \propto t^{-1/2}$. These upper limits are shown in Table \ref{table1} and plotted in Figure \ref{fig:lim}.

\section{Implications of the Upper Limits}

One can use the flux and luminosity upper limits derived above to constrain the parameters of FRB physical models. We consider two different constraints, namely, the magnetar spindown model and the GRB external shock model.

\subsection{Rotation-powered new-born magnetar}

First, we consider the possibility that the central engine of FRB 180814.J0422+73 is a young magnetar, whose spindown luminosity may be partially converted in the LAT energy band.
The spin-down luminosity of a magnetar is \citep{shapiro83}
\begin{equation}
 L_{\text{sd}}=\frac{L_{\text{sd,}0}}{\left( 1+T/T_{\text{sd}}\right) ^{2}},
\end{equation}
where the characteristic spin-down luminosity $L_{\text{sd,}0}$ and the timescale $T_{\text{sd}}$ can be expressed as \citep{2017zhang}
\begin{eqnarray}
 L_{\text{sd,}0}&=&{1.0\times 10^{45}}\text{ erg s}^{-1}B_{p,13}^{2}P_{0,-3}^{-4}R_{6}^{6},\\
 T_{\text{sd}}&=&0.65\text{ yr }I_{45}B_{p,13}^{-2}P_{0,-3}^{2}R_{6}^{-6}.
\end{eqnarray}
Here, $B_p$, $P_0$, $R$ and $I$ are the polar magnetic dipole field strength on the surface, the initial rotation period, the radius, and the moment of inertia of the magnetar. As a convention, we employ $Q=10^nQ_n$ in cgs units. The corresponding $\gamma$-ray luminosity can be written as
\begin{equation}\label{eq:etaL}
 L_{\gamma }=\eta L_{\text{sd}},
\end{equation}
where $\eta =\eta _{\gamma }f_{b}^{-1}$ is the efficiency parameter, with $f_{b}$ being the beaming factor and $\eta _{\gamma } < 1$ being ratio between the radiated isotropic equivalent $\gamma $-ray luminosity and the spin-down luminosity of the magnetar.
From observational constraints, we require
\begin{equation}\label{eq:lim}
 L_{\gamma }\left( T\right) \leq L_{\lim }\left( T\right),
\end{equation}
where $L_{\lim }\left( T\right) =L_{\lim }\left( 10\text{ yr}/T\right) ^{1/2}$, and $L_{\lim }\simeq {7.3\times 10^{43}}$~erg~s$^{-1}$ is the observed 10-year upper limit of luminosity, which is presented in Table \ref{table1}. By substituting Equation (\ref{eq:etaL}) into Equation (\ref{eq:lim}), the allowed parameter space of the initial rotation period $P_0$ and the surface magnetic field strength $B_p$ of the magnetar can be calculated.

Next comes the discussion about the age, $T$, of the magnetar. If a magnetar is generated by the death of a massive star, it should be initially surrounded by thick supernova ejecta. The transparency time $T_{\text{trans}}$ corresponds to the epoch when gamma-ray radiation becomes transparent to the ejecta. On the other hand, if a magnetar is generated by the merger of two neutron stars \citep{dai06, fan06, metzger08, zhang13}, it would not necessarily have a heavy shell, and the transparency time could be short, e.g., $T_{\text{trans}}\sim 10^3$ s \citep{sun17}. Generally speaking, the transparency time depends on the mass and the opacity of the ejecta, which lasts from minutes to years.
If the magnetar was born before January 1, 2009, and the gamma rays had been already transparent when the observations started, the age $T$ can be greater than ten years. If the magnetar was born after January 1, 2009, we can set $T = T_{\text{trans}}$, which gives the tightest constraint. Because the upper limits vary slightly in different half-year time bins, we use the average upper limit for simplicity.

\begin{figure}[tbph]
\begin{center}
\includegraphics[width=0.65\textwidth,angle=0]{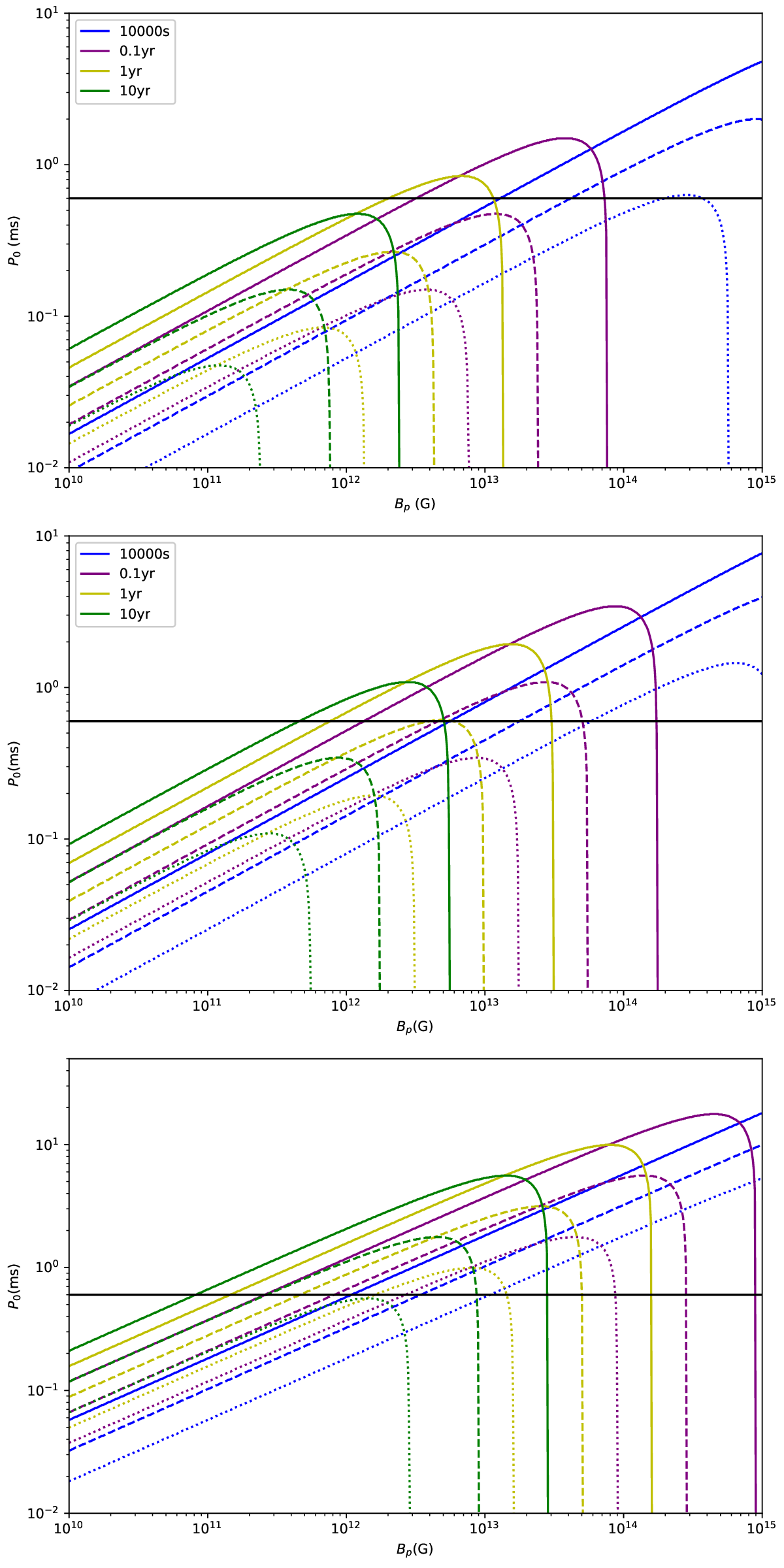}
\end{center}
\caption{ {Constraints of the initial spin period and the surface magnetic field of a putative magnetar engine from the LAT upper limit.} Dotted lines, dashed lines and solid lines represent cases of $\eta  =0.01$, 0.1 and 1, respectively. The allowed parameter space is above each curve for each case. The horizontal line indicates the lower limit of the initial period of a magnetar is $P_0 = 0.6$~ms. The redshift of FRB 180814.J0422+73 in the top, middle and bottom panel is assumed to be $z=0.11$, 0.05 and 0.01, respectively.
}
\label{fig:BPlim}
\end{figure}

We consider a range of different ages of the magnetar, i.e., $T=10^4$ s to 10~yr, to perform the constraints. For each $T$, three efficiency values, $\eta =0.01$, 0.1, and 1, are used. The corresponding constraints on the magnetar parameters are shown in the top panel of Figure \ref{fig:BPlim}. The allowed parameter space is the region above each corresponding curve for a given value of $T$ and $\eta$. It is obvious that the allowed range is tighter for a higher efficiency or a shorter age. The initial rotation period of a magnetar should satisfy $P_0\gtrsim 0.6$~ms \citep[e.g.][]{vink06}, which is marked as a horizontal line in Figure \ref{fig:BPlim}. For high efficiencies, e.g., $\eta=1$, the magnetar parameter space is constrained if the age is shorter than one year. The source has to spin relatively slowly if it is a magnetar. For lower efficiencies (e.g. $\eta = 0.01$), there is essentially no constraint. Since $z\sim 0.11$ is only the upper limit of the redshift for FRB 180814.J0422+73, we additionally check how the magnetar parameters are constrained if FRB 180814.J0422+73 is closer by. By assuming two smaller redshift values, we plot such constraints in the middle and bottom panel of Figure \ref{fig:BPlim}. As expected, the smaller the redshift, the tighter the allowed parameter ranges. In other words, a longer initial rotation period and/or a smaller surface magnetic field are required if the source is nearby.

\subsection{GRB blastwave}

In view of the possibility of a long GRB as the progenitor of the FRB source \citep[e.g.][]{megz17}, we apply the upper limits to constrain the existence of such a GRB in the direction of FRB 180814.J0422+73 during the past ten years. A similar constraint was made by \cite{xi17} for a possible association between FRB 131104 and a putative GRB \citep{delaunay16}.

If there was a long GRB beaming towards Earth during the time span of the LAT observation, high-energy synchrotron radiation can be generated by the accelerated electrons from the external forward shock. We calculate the brightness of GeV afterglow within the framework of the standard GRB afterglow model (\citealt{gao13} and references therein) and use the upper limits to constrain the model parameters, especially the isotropic energy of the blastwave.

We consider a blastwave with total isotropic energy $E_{\text{iso}}$, initial Lorentz factor $\Gamma_0$, initial thickness $\Delta_0$ and jet opening angle $\theta _j$ running into a circumburst medium (CBM). The thermal energy in the shock is deposited proportionally into electrons and magnetic fields in fraction of $\varepsilon _e$ and $\varepsilon _B$, respectively.
The CBM has a density profile $n\left( R \right)=A R^{-k}, 0 < k \leq 4$ \citep{bland76}, where $R$ is the distance from the central engine. We consider two commonly used density profiles, i.e. a constant density interstellar medium (ISM) ($k = 0$ and $A=n_0$) and a free stratified wind ($k = 2$). For the wind model, one has
\begin{eqnarray}
 A&=&\frac{\dot{M}}{4\pi m_p v_{w}}={3\times 10^{35}}\dot{M}_{-5}v_{8}^{-1}\text{ cm}^{-1} \nonumber\\
 &=&{3\times 10^{35}}A_\ast \text{ cm}^{-1},
\end{eqnarray}
where $\dot{M}=10^{-5}\dot{M}_{-5}\text{M}_{\odot }$~yr$^{-1}$, $v_{w}=10^{8}v_8$ cm s$^{-1}$ are the mass loss rate and wind velocity of the progenitor, respectively.

The blastwave enters the Blandford-McKee self-similar deceleration phase \citep{bland76} after the reverse shock crosses the jet. For $z \ll 1$, the deceleration time for an impulsive fireball is \citep{zhang19}
\begin{equation}
 t_{\text{dec}} \simeq
 \left\{
 \begin{array}{ll}
 (185 \text{ s}) \ E_{52}^{1/3}n_0^{-1/3} \Gamma_{0,2}^{-8/3}, & \text{ ISM}, \\
 (0.9 \text{ s}) \ E_{52}A_{\ast,-1}^{-1} \Gamma_{0,2}^{-4}, & \text{ wind.}
 \end{array}
 \right.
\end{equation}
For an on-axis GRB, the GRB afterglow lightcurve peaks at the reverse shock crossing time $t_{\times} =\max \left( t_{\text{dec}},T\right)$, where $T=\Delta _{0}/c$ is the duration of the burst. In the following calculations, we adopt $\Gamma _{0}=100$, $\Delta _{0}=10^{12}$~cm and $\theta _j =0.1$.

\subsubsection{The ISM model}

First, we discuss the on-axis GRB in the ISM model. If the jet beams towards Earth, the flux peaks at the crossing time. If $E_{52}>{5.8\times 10^{-3}}n_{0}$, it corresponds to the thin shell regime. So the reverse shock is Newtonian (NRS), and it crosses the jet at the deceleration time. If $E_{52}\leq {5.8\times 10^{-3}}n_{0}$, it corresponds to the thick shell regime with a relativistic reverse shock (RRS), and the crossing time is equal to $T$ \citep{sari95}.

\begin{figure}
\centering
\includegraphics[width=0.8\textwidth]{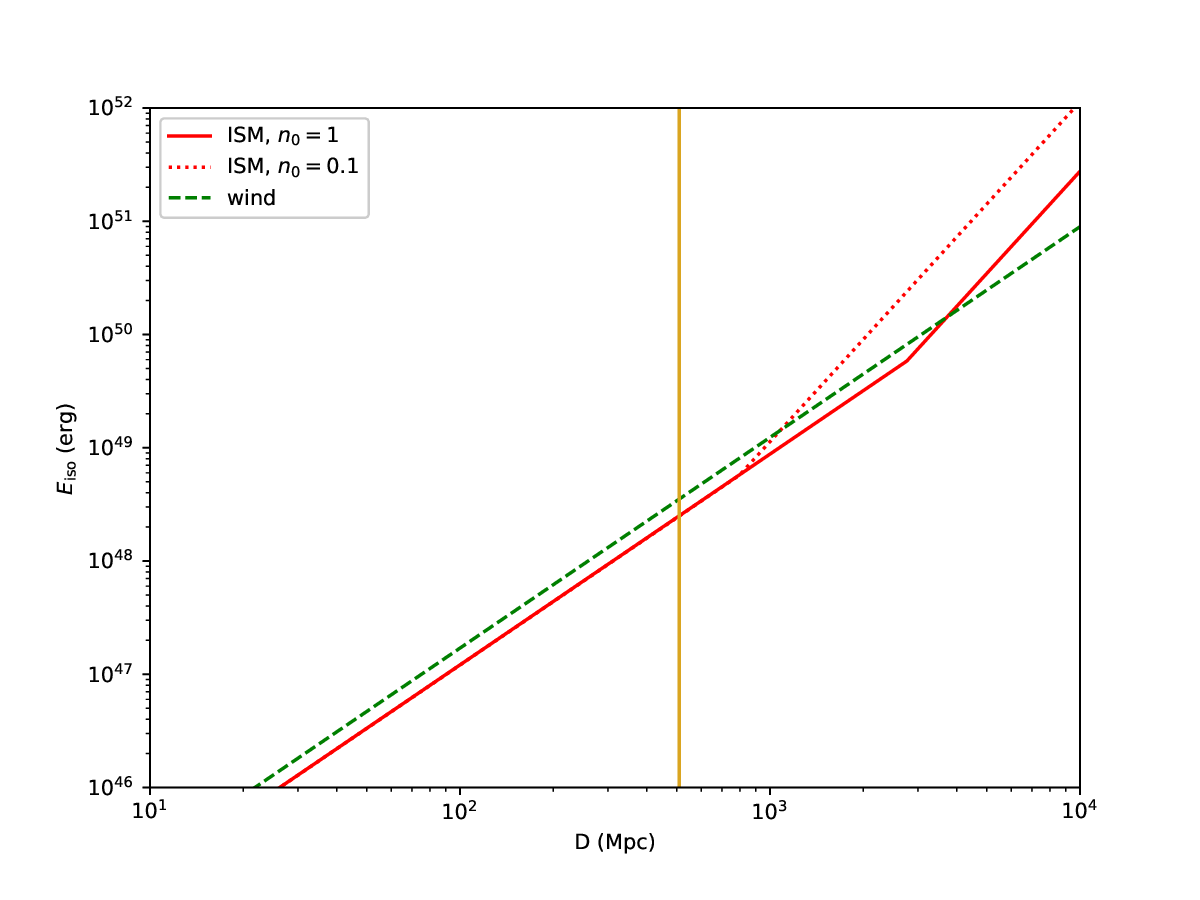}
\caption{Constraints on the isotropic kinetic energy of the blastwave and the luminosity distance of the source for the on-axis GRB external shock model. The solid and dotted lines represent for the upper limits of $E_\text{iso}$ with the ISM number density $n_{0}=1$~cm$^{-3}$ and 0.1~cm$^{-3}$, respectively. The dashed lines are upper limits for the wind model with $A_{\star }=0.1$.
Besides, we set microphysics shock parameters to be $\varepsilon _{e}=0.1$, $\varepsilon _{B}=0.01$ and electron index $p=2.3$.}
\label{fig:EDonlim}
\end{figure}

At the crossing time, the corresponding frequency of the minimum electron Lorentz factor is
\begin{equation}
 \nu _{m} \simeq
 \left\{
 \begin{array}{ll}
 {3.8\times 10^{14}}\text{ Hz }\varepsilon _{e,-1}^{2}\varepsilon _{B,-2}^{1/2}n_{0}^{1/2}, & \text{NRS}, \\
 {5.0\times 10^{15}}\text{ Hz }E_{52}^{1/2}\varepsilon _{e,-1}^{2}\varepsilon _{B,-2}^{1/2}, & \text{RRS}\text{.} \\
 \end{array}
 \right.
\end{equation}
The cooling frequency is
\begin{equation}
 \nu _{c} \simeq \left\{
 \begin{array}{lr}
 {9.5\times 10^{17}}\text{ Hz }E_{52}^{-2/3}\varepsilon _{B,-2}^{-3/2}n_{0}^{-5/6}, & \text{NRS}, \\
 {2.2\times 10^{18}}\text{ Hz }E_{52}^{-1/2}\varepsilon _{B,-2}^{-3/2}n_{0}^{-1}, & \text{ }\text{RRS}\text{.}
 \end{array}
 \right.
\end{equation}
The peak flux density at a luminosity distance of $D$ from the source is
\begin{equation}
 F_{\nu ,\max } \simeq 5.5\text{ mJy }E_{52}\varepsilon _{B,-2}^{1/2}n_{0}^{1/2}D_{28}^{-2},
\end{equation}
which does not depend on $t$ (and hence, has no difference between the thin and thick shell cases).
The $\gamma $-ray band usually satisfies $\nu >$max$(\nu _{m},\nu _{c})$. The flux density at frequency $\nu$ can be calculated as
\begin{equation}
 F_{\nu }=F_{\nu ,\max }\left( \frac{\nu _{c}}{\nu _{m}}\right) ^{-\frac{p-1}{2}}\left( \frac{\nu }{\nu _{c}}\right) ^{-\frac{p}{2}}
\end{equation}
where $p$ is the power-law index of the electron density distribution. The integrated flux in the LAT-band is
\begin{eqnarray}
 F_{\gamma }\left( t_{\times}\right) &=&\int_{100\text{ MeV}}^{10\text{ GeV}}\frac{F_{\nu }}{h}dh\nu \notag \\
 &\simeq &\left\{
 \begin{array}{r}
 {2.4\times 10^{-10}}\text{ erg cm}^{-2}\text{ s}^{-1}\text{ }E_{52}^{2/3}\varepsilon _{e,-1}^{p-1}\varepsilon _{B,-2}^{(p-2)/4} \\
 \times n_{0}^{(3p-2)/12}D_{28}^{-2}, \qquad \text{NRS}, \\
 {1.9\times 10^{-9}}\text{ erg cm}^{-2}\text{ s}^{-1}\text{ }E_{52}^{(p+2)/4} \varepsilon _{e,-1}^{p-1} \varepsilon _{B,-2}^{(p-2)/4}\\
 \times D_{28}^{-2}, \qquad \text{RRS}\text{,}
\end{array}
\right. \notag \\
&&
\end{eqnarray}
where the constant coefficients (as well the constants in the following formulae of the integrated flux for the wind model) are obtained by substituting $p=2.3$. The peak flux should satisfy
\begin{equation}\label{eq:Flim}
 F_{\gamma }\leq F_{\lim }\simeq {1.1\times 10^{-11}}\text{ erg cm}^{-2}\text{s}^{-1}
\end{equation}
We can then constrain the isotropic energy by
\begin{equation}
 E_{52}\lesssim \left\{
 \begin{array}{lr}
 {9.4\times 10^{-3}}\varepsilon _{e,-1}^{3(1-p)/2}\varepsilon _{B,-2}^{3(2-p)/8}n_{0}^{(2-3p)/8}D_{28}^{3}, &\text{ }\text{NRS}, \\
 {7.8\times 10^{-3}}\varepsilon _{e,-1}^{4(1-p)/(p+2)}\varepsilon _{B,-2}^{(2-p)/(p+2)}D_{28}^{8/(p+2)}, &\text{RRS}\text{.}
 \end{array}
 \right.
\end{equation}
We plot this limit in Figure \ref{fig:EDonlim}.
The density of ISM has no effect on the limiting result below the luminosity distance upper limit. The isotropic energy is constrained as $E_\text{iso}\lesssim 10^{48}$~erg for the luminosity distance below $D\sim 500$~Mpc.

\subsubsection{The wind model}

The same exercise can be applied to the wind model. If the jet beams towards Earth, the flux peaks at the crossing time. The case for $E_{52}>37A_{\ast ,-1}^{-1}$ corresponds to a Newtonian reverse shock, and $t_\times=t_{\text{dec}}$. Otherwise, the crossing time is $T$, a relativistic reverse shock is expected.
At $t_\times$, the minimum injection frequency is
\begin{equation}
 \nu _{m} \simeq
 \left\{
 \begin{array}{cc}
 {5.8\times 10^{17}}\text{ Hz }E_{52}^{-1}\varepsilon _{e,-1}^{2}\varepsilon _{B,-2}^{1/2}A_{\ast ,-1}^{3/2}, & \text{NRS}, \\
 {2.6\times 10^{15}}\text{ Hz }E_{52}^{1/2}\varepsilon _{e,-1}^{2}\varepsilon _{B,-2}^{1/2}, & \text{RRS}.
 \end{array}
 \right.
\end{equation}
The cooling frequency is
\begin{equation}
 \nu _{c} \simeq \left\{
 \begin{array}{r}
 {1.5\times 10^{16}}\text{ Hz }E_{52}\varepsilon _{B,-2}^{-3/2}A_{\ast ,-1}^{-5/2}, \text{ }\text{NRS}, \\
 {9.3\times 10^{16}}\text{ Hz }E_{52}^{1/2}\varepsilon _{B,-2}^{-3/2}A_{\ast ,-1}^{-2}, \text{RRS}\text{.}
 \end{array}
 \right.
\end{equation}
The peak flux density is
\begin{equation}
 F_{\nu ,\max } \simeq \left\{
 \begin{array}{lr}
 180\text{ mJy }\varepsilon _{B,-2}^{1/2}A_{\ast ,-1}^{3/2}D_{28}^{-2}, & \text{NRS}, \\
 29\text{ mJy }E_{52}^{1/2}\varepsilon _{B,-2}^{1/2}A_{\ast ,-1}D_{28}^{-2}, & \text{RRS}\text{.}
 \end{array}
 \right.
\end{equation}
The integrated $\gamma$-ray flux in LAT-band is
\begin{eqnarray}\label{eq:F2}
 F_{\gamma }\left( t_{\times}\right) \simeq \left\{
 \begin{array}{rr}
 {1.1\times 10^{-7}}\text{ erg cm}^{-2}\text{ s}^{-1}\text{ }E_{52}^{\left(2-p\right) /2}\varepsilon _{e,-1}^{p-1}\varepsilon _{B,-2}^{(p-2)/4} \\
 \times A_{\ast ,-1}^{(3p-2)/4}D_{28}^{-2}, \qquad \text{NRS}, \\
 {1.3\times 10^{-9}}\text{ erg cm}^{-2}\text{ s}^{-1}\text{ }E_{52}^{(p+2)/4} \varepsilon _{e,-1}^{p-1} \varepsilon _{B,-2}^{(p-2)/4}\\
 \times D_{28}^{-2}, \qquad \text{RRS}\text{.}
 \end{array}
\right. \notag \\
&&
\end{eqnarray}

By substituting Equation (\ref{eq:F2}) into Equation (\ref{eq:Flim}), the limit of isotropic energy can be obtained for $p>2$
\begin{eqnarray}\label{eq:limE2}
\left\{
\begin{array}{ll}
	 E_{52}&\gtrsim  {7.1\times 10^{26}}\varepsilon _{e,-1}^{2\left( p-1\right) /\left( p-2\right)}\varepsilon _{B,-2}^{1/2}A_{\ast ,-1}^{(3p-2)/2\left( p-2\right)} D_{28}^{4/\left( 2-p\right) }\\
 	&  \qquad \qquad \qquad \qquad \qquad \qquad \qquad \qquad \text{NRS   (a);} \\
 	E_{52} &\lesssim  {1.1\times 10^{-2}}\varepsilon _{e,-1}^{4(1-p)/(p+2)}\varepsilon _{B,-2}^{(2-p)/(p+2)}D_{28}^{8/(p+2)}\\
 	&  \qquad \qquad \qquad \qquad \qquad \qquad \qquad \qquad \text{RRS   (b)}\text{.}\\
 	\end{array}
\right.\notag \\
&
\end{eqnarray}
Since for the typical parameters one has $t_{\rm dec}  \ll \Delta _0/c$ for the wind model, the NRS case is not relevant. The limit constrained by Equation (\ref{eq:limE2}b) is plotted in Figure \ref{fig:EDonlim} with the same parameters as the previous model except for $A_{\star }=0.1$.

\subsection{Orphan afterglow}

A magnetar central engine of repeating FRBs can be in principle produced by a GRB not beaming towards Earth. If this is the case, then there could be an ``orphan'' afterglow associated with the unseen GRB \citep{granot02}. Our upper limit can also pose some constraints on the parameters if the viewing angle is not too much larger than the jet opening angle.

In general, the bulk Lorentz factor of the blastwave in the self-similar deceleration phase is given by \citep{gao13}
\begin{equation}
 \Gamma =\left[ \frac{\left( 17-4k\right) E_{\text{iso}}}{4^{5-k}\left(4-k\right) ^{3-k}\pi Am_{p}c^{5-k}t^{3-k}}\right] ^{\frac{1}{2\left(4-k\right) }}
\end{equation}
where $m_p$ is the mass of proton and $c$ is the speed of light.
The flux of high-energy synchrotron radiation generated at jet-break time $t_{j}$ is \citep{kumar09}
\begin{equation}\label{eq:F3}
 F_{\gamma }\left( t_{j}\right) =F_{\gamma }\left( t_{\times}\right) \left( \frac{t_{j}}{t_{\times}}\right) ^{\frac{2-3p}{4}}, \notag
\end{equation}
which corresponds to the epoch when $\Gamma$ is decelerated to $1/\theta_j$. Assuming that the sideways expansion effect is negligible, one can calculate the fluxes when $\Gamma$ is decelerated to $1/ \theta_v$, which roughly corresponds to the peak flux of an observer with viewing angle $\theta_{v}$. The flux reduction factor $\Gamma ^2 \theta _j ^2$ due to the edge effect is considered. It means that $F_{\gamma }\left( t>t_{j}\right)\propto t^{-(3p+1)/4} (t^{-3p/4}) $ for the ISM (wind) model, thus the viewing angle should be greater than $1.6\theta _j $ ($1.1\theta _j $) by assuming $E_{\rm iso}=10^{52}$~erg. If we take a higher shock energy $E_{\rm iso}=10^{55}$~erg into consideration, a larger viewing angle $\theta _v > 3.9\theta _j $ must be satisfied for the ISM model. Meanwhile, there is no constraint of $\theta _v$ for the wind model under the luminosity distance upper limit.

\section{Summary and Discussions}

We have derived the flux and luminosity upper limits in the 100~MeV$-$10~GeV energy range on the FRB 180814.J0422+73 source using the ten-year {\it Fermi}/LAT data from 2009 January 1 to 2018 December 29. The average flux upper limit of the 20 six-month time bins is $\sim {1.1\times 10^{-11}}$~erg~cm$^{-2}$~s$^{-1}$. Consider the maximum redshift is $z=0.11$, then the upper limit of the luminosity is $\sim {3.3\times 10^{44}}$~erg~s$^{-1}$. The ten-year upper limits of flux and luminosity are $\sim {2.4\times 10^{-12}}$~erg~cm$^{-2}$~s$^{-1}$ and $\sim {7.3\times 10^{43}}$~erg~s$^{-1}$, respectively.

These upper limits can be used to constrain the model parameters of some FRB progenitor models. For a young magnetar central engine, if the efficiency parameter $\eta$ is large (which means that the beaming effect is extremely strong), the magnetar parameters can be constrained, in particular, the magnetar cannot spin too fast. If $\eta$ is small or if the transparency time for a new-born magnetar is too long, then there is virtually no constraint. One can also constrain the existence of a GRB in the direction of FRB 180814.J0422+73 during the ten-year span of LAT observations. An on-beam typical long GRB is ruled out by the upper limits regardless of whether the circumburst medium is an ISM or a wind.
For the orphan afterglow case, we constrain that the viewing angle has to be much greater than the jet opening angle for the ISM model with a typical kinetic energy. No constraint can be derived for the wind model.

These conservative constraints are derived by assuming a maximum redshift 0.11 for FRB 180814.J0422+73. The true redshift could be smaller. If the host galaxy of FRB 180814.J0422+73 is determined in the future and it turns out that it is much smaller than 0.11, much tighter constraints can be obtained.

\acknowledgments

We thank Liangduan Liu, Zongkai Peng for discussion and Hezhen Liu, Li Feng for helpful information. BBZ acknowledges support from National Thousand Young Talents program of China and National Key Research and Development Program of China (2018YFA0404204) and The National Natural Science Foundation of China (Grant No. 11833003). 

\clearpage

\begin{table}
\label{table1}
\centering
\caption{Upper Limits of LAT Observations of FRB 180814.J0422+73}
\label{table:1}
\begin{center}
\begin{tabular}{llccc}
\toprule
t1& t2& Photon Flux& Energy Flux& Luminosity
\begin{tablenote}
 {The luminosity upper limits are calculated on account of the redshift $z=0.11$.}
\end{tablenote}\\
 & & (ph cm$^{-2}$~s$^{-1}$)& (erg cm$^{-2}$~s$^{-1}$)& (erg s$^{-1}$)\\
\hline
2009-01-01&2009-07-02& ${1.7\times 10^{-8}}$& ${1.3\times 10^{-11}}$& ${4.0\times 10^{44}}$\\
2009-07-02&2010-01-01& ${2.1\times 10^{-8}}$& ${1.6\times 10^{-11}}$& ${4.9\times 10^{44}}$\\
2010-01-01&2010-07-02& ${1.1\times 10^{-8}}$& ${8.3\times 10^{-12}}$& ${2.6\times 10^{44}}$\\
2010-07-02&2011-01-01& ${1.2\times 10^{-8}}$& ${8.9\times 10^{-12}}$& ${2.8\times 10^{44}}$\\
2011-01-01&2011-07-02& ${1.6\times 10^{-8}}$& ${1.2\times 10^{-11}}$& ${3.7\times 10^{44}}$\\
2011-07-02&2012-01-01& ${1.3\times 10^{-8}}$& ${9.8\times 10^{-12}}$& ${3.0\times 10^{44}}$\\
2012-01-01&2012-07-01& ${1.4\times 10^{-8}}$& ${1.1\times 10^{-11}}$& ${3.3\times 10^{44}}$\\
2012-07-01&2012-12-31& ${1.1\times 10^{-8}}$& ${8.1\times 10^{-12}}$& ${2.5\times 10^{44}}$\\
2012-12-31&2013-07-01& ${1.8\times 10^{-8}}$& ${1.4\times 10^{-11}}$& ${4.2\times 10^{44}}$\\
2013-07-01&2013-12-31& ${2.0\times 10^{-8}}$& ${1.5\times 10^{-11}}$& ${4.7\times 10^{44}}$\\
2013-12-31&2014-07-01& ${1.0\times 10^{-8}}$& ${7.7\times 10^{-12}}$& ${2.4\times 10^{44}}$\\
2014-07-01&2014-12-31& ${1.1\times 10^{-8}}$& ${8.1\times 10^{-12}}$& ${2.5\times 10^{44}}$\\
2014-12-31&2015-07-01& ${1.2\times 10^{-8}}$& ${8.9\times 10^{-12}}$& ${2.8\times 10^{44}}$\\
2015-07-01&2015-12-30& ${8.9\times 10^{-9}}$& ${6.7\times 10^{-12}}$& ${2.1\times 10^{44}}$\\
2015-12-30&2016-06-30& ${1.4\times 10^{-8}}$& ${1.0\times 10^{-11}}$& ${3.2\times 10^{44}}$\\
2016-06-30&2016-12-29& ${1.1\times 10^{-8}}$& ${8.3\times 10^{-12}}$& ${2.6\times 10^{44}}$\\
2016-12-29&2017-06-30& ${1.7\times 10^{-8}}$& ${1.2\times 10^{-11}}$& ${3.9\times 10^{44}}$\\
2017-06-30&2017-12-29& ${1.5\times 10^{-8}}$& ${1.1\times 10^{-11}}$& ${3.5\times 10^{44}}$\\
2017-12-29&2018-06-30& ${1.2\times 10^{-8}}$& ${9.0\times 10^{-12}}$& ${2.8\times 10^{44}}$\\
2018-06-30&2018-12-29& ${1.7\times 10^{-8}}$& ${1.3\times 10^{-11}}$& ${3.4\times 10^{44}}$\\

\hline
Average Value\begin{tablenote}
 {These are the average of all the half-year upper limits above.}
\end{tablenote}
& & ${1.4\times 10^{-8}}$& ${1.1\times 10^{-11}}$& ${3.3\times 10^{44}}$\\
2009-01-01&2018-12-29& ${3.2\times 10^{-9}}$& ${2.4\times 10^{-12}}$& ${7.3\times 10^{43}}$\\
2018-08-14&2018-12-29& ${2.0\times 10^{-8}}$& ${1.5\times 10^{-11}}$& ${4.6\times 10^{44}}$\\
\hline
\end{tabular}
\end{center}

\end{table}

\begin{thebibliography}

\bibitem[Abdo et al.(2010a)]{abdo10a} Abdo, A.~A., Ackermann, M., Ajello, M., et al.\ 2010a, \aap, 524, A75
\bibitem[Abdo et al.(2010b)]{abdo10b} Abdo, A.~A., Ackermann, M., Ajello, M., et al.\ 2010b, \apjs, 188, 405
\bibitem[CHIME/FRB Collaboration et al.(2019)]{2019chime} Amiri, M., Bandura, K., et al. CHIME/FRB Collaboration, \ 2019, \nat, 566, 235

\bibitem[Blandford \& McKee(1976)]{bland76} Blandford, R.~D., \& McKee, C.~F.\ 1976, Physics of Fluids, 19, 1130

\bibitem[Caleb et al.(2019)]{caleb19} Caleb, M., Stappers, B.~W., Rajwade, K., \& Flynn, C.\ 2019, \mnras, 484, 5500
\bibitem[Champion et al.(2016)]{cham16} Champion, D.~J., Petroff, E., Kramer, M., et al.\ 2016, \mnras, 460, L30
\bibitem[Chatterjee et al.(2017)]{chat17} Chatterjee, S., Law, C.~J., Wharton, R.~S., et al.\ 2017, \nat, 541, 58

\bibitem[Dai et al.(2006)]{dai06} Dai, Z.~G., Wang, X.~Y., Wu, X.~F., \& Zhang, B.\ 2006, Science, 311, 1127
\bibitem[DeLaunay et al.(2016)]{delaunay16} DeLaunay, J.~J., Fox, D.~B., Murase, K., et al.\ 2016, \apjl, 832, L1

\bibitem[Fan \& Xu(2006)]{fan06} Fan, Y.-Z., \& Xu, D.\ 2006, \mnras, 372, L19
\bibitem[Feldman \& Cousins(1998)]{feldman98} Feldman, G.~J., \& Cousins, R.~D.\ 1998, \prd, 57, 3873

\bibitem[Gajjar et al.(2018)]{gajj18} Gajjar, V., Siemion, A.~P.~V., Price, D.~C., et al.\ 2018, \apj, 863, 2
\bibitem[Gao et al.(2013)]{gao13} Gao, H., Lei, W.-H., Zou, Y.-C., Wu, X.-F., \& Zhang, B.\ 2013, New Astron. Rev., 57, 141
\bibitem[Granot et al.(2002)]{granot02} Granot, J., Panaitescu, A., Kumar, P., \& Woosley, S.~E.\ 2002, \apjl, 570, L61

\bibitem[Kashiyama \& Murase(2017)]{kashi17} Kashiyama, K., \& Murase, K.\ 2017, \apjl, 839, L3
\bibitem[Keane et al.(2012)]{kean12} Keane, E.~F., Stappers, B.~W., Kramer, M., \& Lyne, A.~G.\ 2012, \mnras, 425, L71
\bibitem[Kumar \& Barniol Duran(2009)]{kumar09} Kumar, P., \& Barniol Duran, R.\ 2009, \mnras, 400, L7

\bibitem[Lorimer et al.(2007)]{lorim07} Lorimer, D.~R., Bailes, M., McLaughlin, M.~A., Narkevic, D.~J., \& Crawford, F.\ 2007, Science, 318, 777
\bibitem[Lyutikov et al.(2016)]{lyu16} Lyutikov, M., Burzawa, L., \& Popov, S.~B.\ 2016, \mnras, 462, 941

\bibitem[Marcote et al.(2017)]{marc17} Marcote, B., Paragi, Z., Hessels, J.~W.~T., et al.\ 2017, \apjl, 834, L8
\bibitem[Margalit et al.(2018)]{margalit18} Margalit, B., Metzger, B.~D., Berger, E., et al.\ 2018, \mnras, 481, 2407
\bibitem[Masui et al.(2015)]{masui15} Masui, K., Lin, H.-H., Sievers, J., et al.\ 2015, \nat, 528, 523
\bibitem[Metzger et al.(2017)]{megz17} Metzger, B.~D., Berger, E., \& Margalit, B.\ 2017, \apj, 841, 14
\bibitem[Metzger et al.(2008)]{metzger08} Metzger, B.~D., Piro, A.~L., \& Quataert, E.\ 2008, \mnras, 390, 781
\bibitem[Murase et al.(2016)]{murase16} Murase, K., Kashiyama, K., \& M{\'e}sz{\'a}ros, P.\ 2016, \mnras, 461, 1498

\bibitem[Nicholl et al.(2017)]{nicholl17} Nicholl, M., Williams, P.~K.~G., Berger, E., et al.\ 2017, \apj, 843, 84

\bibitem[Palaniswamy et al.(2018)]{palaniswamy18} Palaniswamy, D., Li, Y., \& Zhang, B.\ 2018, \apjl, 854, L12
\bibitem[Petroff et al.(2016)]{petr16} Petroff, E., Barr, E.~D., Jameson, A., et al.\ 2016, PASA, 33, e045

\bibitem[Ravi et al.(2015)]{ravi15} Ravi, V., Shannon, R.~M., \& Jameson, A.\ 2015, \apjl, 799, L5
\bibitem[Ravi et al.(2016)]{ravi16} Ravi, V., Shannon, R.~M., Bailes, M., et al.\ 2016, Science, 354, 1249
\bibitem[Roe \& Woodroofe(1999)]{roe99} Roe, B.~P., \& Woodroofe, M.~B.\ 1999, \prd, 60, 053009

\bibitem[Sari \& Piran(1995)]{sari95} Sari, R., \& Piran, T.\ 1995, \apjl, 455, L143
\bibitem[Scholz et al.(2016)]{scho16} Scholz, P., Spitler, L.~G., Hessels, J.~W.~T., et al.\ 2016, \apj, 833, 177
\bibitem[Shapiro \& Teukolsky(1983)]{shapiro83} Shapiro, S.~L., \& Teukolsky, S.~A.\ 1983, Black Holes, White Dwarfs, and Neutron Stars: The Physics of Compact Objects (New York: Wiley)
\bibitem[Spitler et al.(2014)]{spit14} Spitler, L.~G., Cordes, J.~M., Hessels, J.~W.~T., et al.\ 2014, \apj, 790, 101
\bibitem[Spitler et al.(2016)]{spit16} Spitler, L.~G., Scholz, P., Hessels, J.~W.~T., et al.\ 2016, \nat, 531, 202
\bibitem[Sun et al.(2017)]{sun17} Sun, H., Zhang, B., \& Gao, H.\ 2017, \apj, 835, 7

\bibitem[Tendulkar et al.(2017)]{tend17} Tendulkar, S.~P., Bassa, C.~G., Cordes, J.~M., et al.\ 2017, \apjl, 834, L7
\bibitem[Thornton et al.(2013)]{thor13} Thornton, D., Stappers, B., Bailes, M., et al.\ 2013, Science, 341, 53

\bibitem[Vink \& Kuiper(2006)]{vink06} Vink, J., \& Kuiper, L.\ 2006, \mnras, 370, L14

\bibitem[Xi et al.(2017)]{xi17} Xi, S.-Q., Tam, P.-H.~T., Peng, F.-K., \& Wang, X.-Y.\ 2017, \apjl, 842, L8

\bibitem[Zhang(2013)]{zhang13} Zhang, B.\ 2013, \apjl, 763, L22
\bibitem[Zhang(2014)]{zhang14} Zhang, B.\ 2014, \apjl, 780, L21
\bibitem[Zhang(2018)]{zhang19} Zhang, B.\ 2018, The Physics of Gamma-Ray Bursts by Bing Zhang.~ISBN: 978-1-139-22653-0.~Cambridge Univeristy Press, 2018,

\bibitem[Zhang \& Zhang(2017)]{2017zhang} Zhang, B.-B., \& Zhang, B.\ 2017, \apjl, 843, L13
\bibitem[Zhang et al.(2018)]{zhang18} Zhang, Y.~G., Gajjar, V., Foster, G., et al.\ 2018, \apj, 866, 149

\end{thebibliography}
\end{document}